\def\cm{{\rm\thinspace cm}}

\def\erg{{\rm\thinspace erg}}

\def\keV{{\rm\thinspace keV}}
\def\km{{\rm\thinspace km}}

\def\Mpc{{\rm\thinspace Mpc}}
\def\Msun{\hbox{$\rm\thinspace M_{\odot}$}}

\def\s{{\rm\thinspace s}}
\def\yr{{\rm\thinspace yr}}

\def\ergpcmsqps{\hbox{$\erg\cm^{-2}\s^{-1}\,$}}

\def\ergps{\hbox{$\erg\s^{-1}\,$}}

\def\kmps{\hbox{$\km\s^{-1}\,$}}

\def\psqcm{\hbox{$\cm^{-2}\,$}}

\documentstyle[11pt,newpasp,twoside,epsf]{article}
\def\edcomment#1{\iffalse\marginpar{\raggedright\sl#1\/}\else\relax\fi}
\reversemarginpar

\begin{document}
\title{Deep Sky Objects and the X-ray Background}
 \author{A.C. Fabian}
\affil{Institute of Astronomy, Madingley Road, Cambridge CB3 0HA, UK }

\begin{abstract}
Deep observations of the X-ray Sky have now resolved most of the X-ray
Background. Obscured active galaxies are a key component. The obscured
objects which dominate at fluxes above $10^{-15}\ergpcmsqps$ are
discussed and models in which the growth of wind-blowing active nuclei
determine both black hole and galaxy mass are presented. The power
from accretion is large and can easily influence the gas in a host
galaxy, as well as heat the intergalactic medium.
\end{abstract}

\section{Introduction}
Observations with Chandra and XMM are revolutionizing our
understanding of the X-ray Background (XRB). Chandra has at last
resolved the 2--7~keV part of the hard XRB (Mushotzky et al 2000;
Garmire et al 2000; Giacconi et al 2000). Since the 2--7~keV spectrum
of the Background attaches smoothly onto the rest of the spectrum up
to its $\nu F_{\nu}$ peak at about 30~keV, we may fairly say that the
X-ray Background is now beginning to be understood. The soft X-ray
Background was of course resolved by ROSAT (Hasinger et al 1998), but
the spectrum below 1~keV is much steeper than that of the harder
background and dominated by different components. 

In the 2--7~keV band the sources at flux levels of
$\sim10^{-15}\ergpcmsqps$ appear to be roughly one third normal
quasars, one third galaxies and the remaining third are optically
faint (Mushotzky et al 2000; Barger et al 2000). It is plausible that
all are powered by active galactic nuclei (AGN), since the
luminosities are all much greater than for the distributed binaries
etc in galaxies and starbursts (such sources do however appear at
lower flux levels). The nucleus itself must be obscured in the second
two classes of object. Obscured AGN have been a major component of
most models for the XRB since first proposed by Setti \& Woltjer in
1989. The reason is the spectrum of the XRB which is flatter than any
known class of extragalactic X-ray source (see Boldt 1987; Fabian \&
Barcons 1992). A range of absorption and redshift in the sources can
result in the observed spectrum (Madau et al 1994; Matt \& Fabian
1994; Comastri et al 1995; Wilman \& Fabian 1999).

In this talk I shall concentrate on the absorbed sources. First I
briefly review nearby obscured AGN, then summarise Chandra results
which show that the optically faint sources are likely obscured
quasars. Next I show that most accretion power is in distant obscured
AGN and that the obscuring medium must cover most of the Sky subtended
at such nuclei. Finally I show that a model for galaxy formation in
which a significant fraction of the central matter in a galaxy remains
in the form of X-ray absorbing gas which is eventually blown away by a
wind from the central quasar when it is powerful enough, can explain
the development of both black holes in galaxy bulges and the XRB.

\section{Obscured AGN}
\subsection{Seyfert II galaxies}
Obscured AGN, in which the power source lies behind a significant
column density of gas which is local to the nucleus, are common and
generally known as Seyfert II galaxies. Strictly speaking, the name
refers to the optical appearance of the object (ie absence of broad
lines and presence of narrow lines characteristic of photoionized gas)
but I shall use Type II to denote any object where the nucleus is
absorbed (i.e. even so absorbed that no narrow lines from photoionized
gas are detectable). The column densities range from below $10^{22}$
to above $10^{25}\psqcm$. The PDS instrument on BeppoSAX has played an
important role here in probing the higher column density regime (e.g.
Maiolino et al 1998). There appear to be at least as many
Compton-thick Seyfert IIs ($N_{\rm H}>1.5\times 10^{24}\psqcm$) as
Compton thin ones ($N_{\rm H}<1.5 \times 10^{24}\psqcm$).

Just how common Seyfert II galaxies are is uncertain. Optical-based
estimates suggest about 3 times that of Seyfert I galaxies, but they
are often insensitive to the Compton-thick ones with high covering
fraction. Unfortunately there is as yet no survey of the hard X-ray
sky which can sort this out in a definitive way. This is a task for
EXIST (Grindlay 2000). For now we can make use of rough arguments such
as that of Matt et al (2000). We note there that the 3 nearest AGN
with luminosities above $10^{40}\ergps$ -- NGC4945, the Circinus
galaxy and Cen A -- all lie within 4~Mpc and have absorbed nuclei. The
column density for Cen A $\sim 10^{23}\psqcm$ and those for NGC4945
and Circinus are 2 and $4\times 10^{24}\psqcm$ respectively. Two of
the three are therefore Compton thick. NGC4945 only appears to host an
AGN at hard X-ray wavelengths. The X-ray luminosity function for
unabsorbed AGN (Miyaji et al 1998) predicts that there is only a 5 per
cent chance of detecting an object within a radius 4~Mpc. The
probability of 3 or more is $2\times 10^{-5}$! Only if obscured AGN
outnumber unobscured ones by about a factor of 10 to one does the
probability exceed two percent. Even though we are using only 3
objects it is highly likely that Type II objects are common, and
present in about 10 per cent of all $L^*$ galaxies. At least one half
of all local ultraluminous IRAS galaxies (ULIRGs) are then obscured
AGN.

\subsection{Obscured quasars -- Type II quasars?}
If most local AGN hide behind large columns of absorbing material, is
it also true for more distant and powerful ones, like the quasars?
Although it might at first thought seem obvious that the answer has
been known for ages, it is not. There is no clear optically-identified
population resembling a more powerful version of the Seyfert II
galaxies. A few objects began to emerge from ASCA and BeppoSAX hard
imaging, but whether they have quasar luminosity is debatable (Halpern
et al 1999). I suspect that such luminous objects are either
unabsorbed or highly absorbed ($N_{\rm H}\gg10^{22}\psqcm$). Perhaps
the radiation pressure and winds from quasars are powerful enough to
blow away intermediate column density material.

The situation is now changing with Chandra data, which routinely shows
a few powerful obscured AGN per image. A simple example is the
brightest serendipitous source in the field of the lensing cluster
A2390. This hard X-ray source has a photometric redshift of about one,
an intrinsic $N_{\rm H}\sim 10^{23}\psqcm$ and a 2-10~keV luminosity
(unabsorbed) of $3\times 10^{44}\ergps$ making it a Type II quasar.
The optical/near infrared appearance and colours show no sign of the
active nucleus. Such sources are common (Crawford et al 2000; Barger
et al 2000).

It is only in the mid infrared that the central power of the object in
the A2390 field (Fabian et al 2000) is revealed since it is detected
by ISO (Lemonon et al 1998). The emission at 7.5 and 15 microns is far
above that expected from the starlight of the host galaxy and is
consistent with that absorbed from the X-ray emitting nucleus (Wilman
et al 2000). The MIR colours indicate that the reradiating dust is
warm to hot (200 -- 1500~K) so probably close to the nucleus. It
should peak its emission at about 60 microns. The infrared connection
is important in showing that we are dealing with absorption of a
quasar in the X-ray band and not some new class of source with a
`funny' spectrum.

\begin{figure}
\plotone{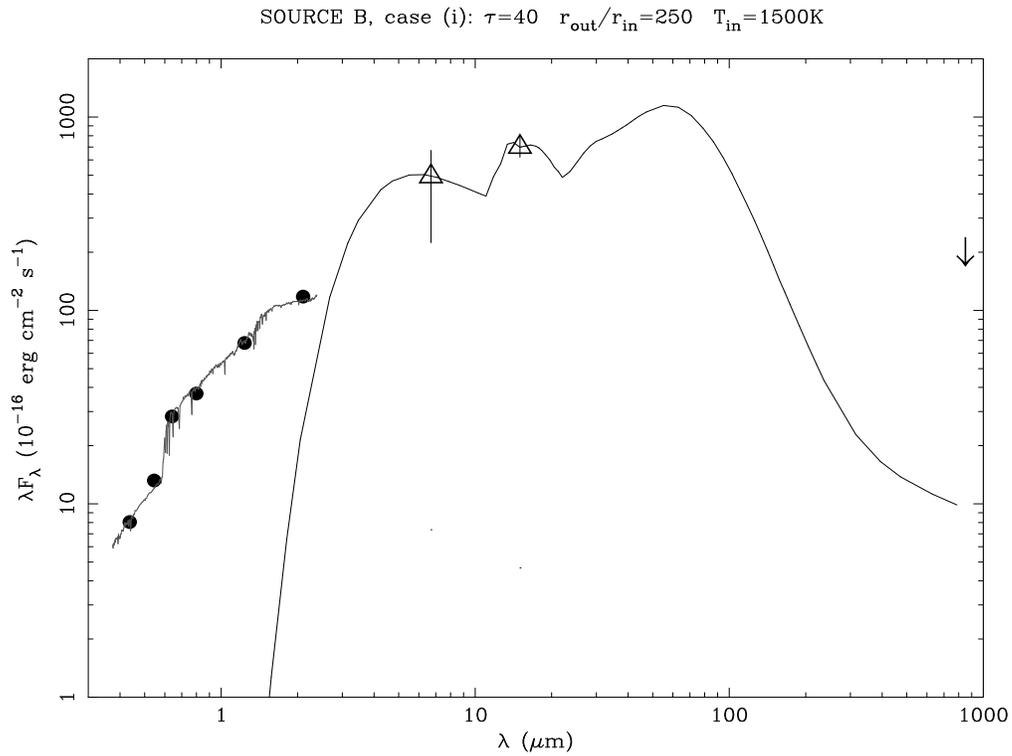}
\caption{Spectral Energy Distribution of lensed X-ray source in the field of
A2390 (Fabian et al 2000). Optical and near-infrared fluxes indicate
an early-type host, ISO mid-infrared points are fitted by a hot dust
model. see Wilman et al 2000. Since this talk was presented, Cowie et
al (2001) have shown that the redshift is significantly greater than
that assumed here ($z\sim 1$).}
\end{figure}

It is also plausible that Type II quasars outnumber normal Type I
quasars. Only then can the spectrum and source counts of the hard XRB
be accounted for. Simple correction of the spectrum of the XRB for
absorption leading to the intrinsic energy density of radiation due to
accretion (Fabian \& Iwasawa 1999) agrees with the local mean density
of massive black holes (Magorrian et al 1998; Merritt \& Ferrarese
2000), if the accretion efficiency is the 10 per cent typical of
radiatively efficient accretion discs. This implies that about 85 per
cent of accretion power has been absorbed (Fabian \& Iwasawa 1999). In
turn this means that the absorbing material covers most of the sky, as
seen by the accreting black holes. To build a massive black hole by
efficient accretion by redshift 2 requires that a quasar-like
luminosity is emitted (Fabian 1999). In conclusion, obscured quasars
must be common.

\section{The central black hole and galaxy formation}
The observed correlation between black hole mass and the stellar mass
of the host bulge (Magorrian et al 1998), and more recently with the
velocity dispersion of that bulge (Gebhardt et al 2000; Ferrarese \&
Merritt 2000), implies some link between the growth of a massive black
hole and the growth of its host galaxy (or at least the bulge of that
galaxy). If the XRB is revealing the growth of massive black holes,
albeit obscured ones, then it should also convey information on the
growth of bulges. Quite how to decipher that information is not yet
clear.

One possibility is that the bulge and black hole grow from gas
together, with the process terminated when a wind from the quasar
becomes powerful enough (Silk \& Rees 1998; Fabian 1999). Consider an
isothermal bulge of velocity dispersion $v$ in which a significant
fraction $f$ of the cooled gas remains in the form of cold dusty
clouds, instead of forming stars. Let the central black hole grow by
accretion and blow a wind of velocity $v_{\rm w}$ with power fraction
$a$ of the Eddington limit, i.e. $L_{\rm w}=a L_{\rm Edd}$. Analogous
to the Eddington limit for point masses, a wind can eject matter from
a distributed mass if the force is large enough (Fabian 1999; Silk \&
Rees 1998 use an energy argument). A column density of gas is ejected
if $$L_{\rm w}>v^4 v_{\rm w} f/G.$$ This happens when the mass of the
hole $$M_{\rm h}\approx{{v^4 \sigma_{\rm T}}\over{4 \pi G^2
m_p}}{v_{\rm w}\over c}{f\over a},$$ which agrees well with the
Gebhardt et al (2000) relation if ${v_{\rm w}\over c}{f\over a}\sim 1$
and the expulsion of the gas also stops the growth of the bulge, as
well as the hole. At the point of expulsion the column density of cold
gas $N\sim \sigma_{\rm T}^{-1}$, so the growth of the hole is Compton
thick.

The expulsion phase can be identified with broad absorption line
quasars, BALQSO, suggesting that $v_{\rm w}\sim 30,000\kmps$. A
normal, unobscured quasar is only seen after gas expulsion and is
presumably due to the draining of the remaining
centrifugally-supported gas close to the hole.

As mentioned in the Introduction, there have been many detailed
attempts to explain the XRB in terms of AGN with various amounts of
absorption over a range of redshifts. Usually the observed present-day
distribution has been extrapolated backwards in time. Richard Wilman,
Paul Nulsen and I (2000) have tried to build a model based on a simple
semi-analytic model for galaxy formation where we work forward in time
from a redshift of about ten. All dark matter haloes initially contain
a seed black hole of mass $M_{\rm seed}$. To make stars the gas
component in the haloes has to cool. Initially the cooling time is
less than the free-fall time throughout the halo producing what we
term a dwarf galaxy (supernova feedback prevents all gas cooling into
stars). Such dwarfs merge until what we term a normal galaxy is built.
This is a galaxy within which the cooling time of its gas exceeds the
free-fall time, so that a hot halo occurs. The central black hole,
which is now many tens of $M_{\rm seed}$ then accretes from that hot
gas (which also progressively cools to make stars) via Bondi
accretion. We assume that the radiative efficiency of accretion is
always 10 per cent and the 2--10~keV luminosity is 3 per cent of the
total. Accretion is terminated by i) exhaustion of the hot gas supply,
ii) a new collapse or iii) the wind from the central quasar exceeding
the above force balance. The last possibility dominates in practice.

If the star formation is inefficient so that $f\sim 0.37$, $M_{\rm
seed}\sim 1.6\times 10^6\Msun$ and there is a final unobscured phase
lasting $9\times 10^7\yr$ we find that we can account for the spectrum
of the XRB, the X-ray luminosity function of quasars and also the
X-ray source counts fairly well (Wilman et al 2000). The model does
however give a spectral excess below 3~keV. This can be fixed by
assuming that gas expulsion is anisotropic, leaving behind a torus of
gas (say that which is centrifugally supported). Otherwise we only
have Compton thick and unobscured objects. We also fail to have many
massive black holes. This is mainly because we do not treat mergers of
normal galaxies and galaxies in groups and clusters.

\begin{figure}
\plotone{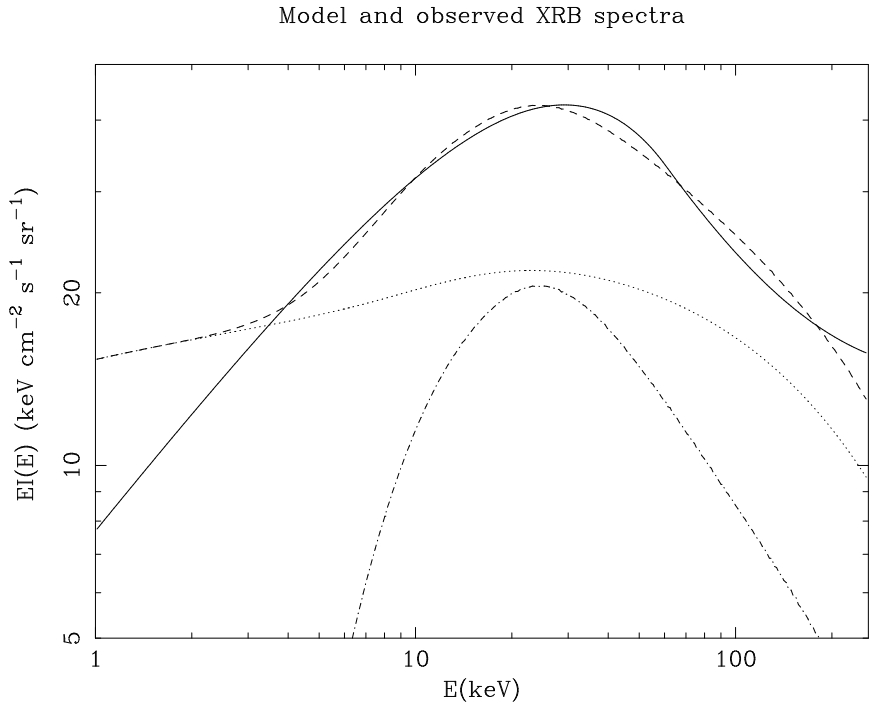}
\caption{XRB spectrum from the model of Wilman et al (2000b). The
solid line indicates the observed spectrum and the dashed line the
model due to the sum of unobscured quasars (dotted line) and
Compton-thick objects (lower dashed line). A modification for
centrifugally-supported gas not ejected by the wind allows the
discrepancy below 3 keV to be eliminated. Note that Chandra has little
sensitivity above 7 keV so should not readily probe the Compton-thick
population in the model (however the model fails to deal with luminous
objects in groups and clusters which may be detected by Chandra if at
high enough redshifts). XMM has better sensitivity up to 10~keV
may pick up the tip of this population. }
\end{figure}
\begin{figure}
\plotone{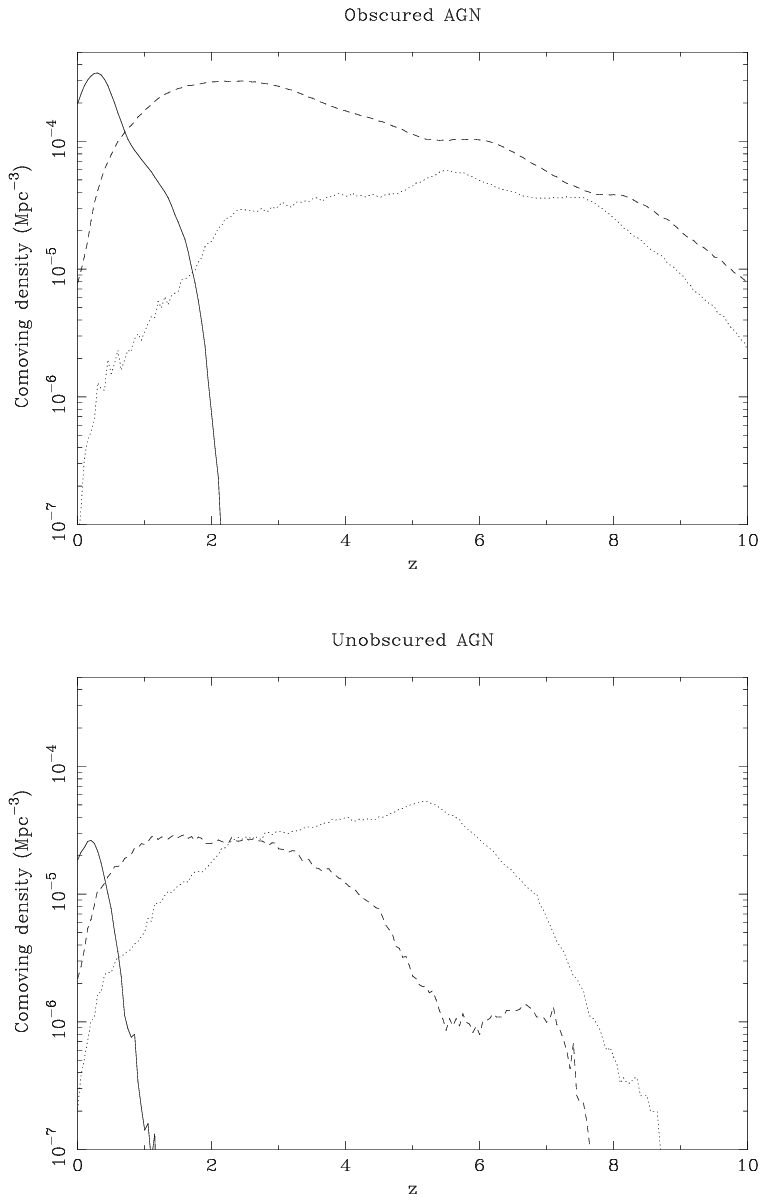}
\caption{The redshift dependence of the X-ray luminosity functions of
the obscured and unobscured populations in the model of Wilman et al
(2000b). Solid lines are objects with intrinsic 2--10~keV luminosities
below $L=10^{43}\ergps$, dashed lines those with
$10^{43}<L<10^{44}\ergps$, and dotted lines those with
$L>10^{44}\ergps$. Note that the model predicts many reasonably
luminous objects above $z$ of 5, although most are obscured.}
\end{figure}

The reasonable success of this model does not of course prove that it
is correct, but it does show that models can be built. We used
accretion of hot gas since it must occur and is relatively easy to
model. It also overcomes the angular momentum problems in the
accretion of cold gas. Other models involving more realistic merger
histories and accretion of cold gas need to be tried. In our model the
black hole plays an important part in the evolution of the galaxy in
that it expels the gas which both fuels the quasar and from which
stars form, thereby ending further star formation.

\section{The power of accretion}

The total power radiated by (obscured) AGN may be 10--50 per cent of
that radiated by stars. If accompanied by winds (as seen in BALQSO)
considerable mechanical energy may result, which is presumably
thermalized in the host galaxy and the intergalactic and intra-group
and -cluster media. Indeed it may be important for shifting the
cluster X-ray luminosity--temperature relation from the gravitational
collapse $L_{\rm x}\propto T^2$ scaling to the observed $T^3$ one (Wu
et al 2000). For ten per cent accretion efficiency, a mean black hole
density of $\rho_{\rm BH}=6\times 10^5\Msun\Mpc^{-3}$ corresponds to
$6\times 10^{58}\erg\Mpc^{-3}$ or $3.7\keV$ particle$^{-1}$ if
$\Omega_{\rm baryon}=0.08, h=0.5$.

The gravitational binding energy of a galactic bulge, where the
velocity dispersion of the bulge is $300 v_{300}\kmps$, is $E_{\rm
bulge}\approx 2\times 10^{-6}v_{300}^2 M_{\rm bulge} c^2$. The energy
from the central black hole $E_{\rm AGN}\approx 5\times 10^{-4}M_{\rm
bulge} c^2$. So only one per cent of that energy can have a major
effect on the formation of that bulge. Of course, the energy needs to
be in the right form. Radiation if ionizing can be damaging, but most
is reradiated quickly. Relativistic jets are probably too fast to
couple to the medium in the bulge well (this could explain why black
holes in radio galaxies grow to be so massive). An uncollimated fast
wind is perhaps the most effective agent.

\section{Final thoughts}
Many deep sky X-ray sources are absorbed AGN. The absorption occurs in
dusty gas and is reradiated in the mid- to far-infrared bands. Chandra
probably will not detect many Compton-thick AGN; XMM may do better if
it can go deep in the band above 7~keV. The presence of the most
massive black holes in the most massive local galactic bulges means
that the most powerful absorbed AGN will be in the most massive
early-type hosts at higher redshifts. The locally observed relation
$M_{\rm BH}\propto v^4$ can be obtained from wind ejection, supporting
evidence that powerful uncollimated winds are common in quasars. Local
Seyfert galaxies may either by younger or rejuvenated black holes,
which may not provide a good guide to the dominant sources of the XRB.
Ultradeep X-ray images may not prove to be the best way to study the
accretion history of the Universe, they may instead reveal more about
the X-ray foreground, i.e. nearby galaxies. Larger numbers of sources
are found by images reaching to the break in the source counts at
$10^{-14}\ergpcmsqps$, the more distant of which may prove to be the
more interesting. There is lots of power due to accretion in the
Universe, most of it obscured and most only directly accessible to
hard X-ray observations.

\end{document}